# Effective Scattering Cross-section in Lattice Thermal Conductivity Calculation with Differential Effective Medium Method

**Di Wu[a)], A. S. Petersen and S. J. Poon[b)]**

*Department of Physics, University of Virginia, Charlottesville, Virginia, 22904-4714*

[a)] Electronic mail: dw7uh@virginia.edu

[b)] Electronic mail: sjp9x@virginia.edu

## ABSTRACT

To further reduce the lattice thermal conductivity of thermoelectic materials, the technique of embedding nanoinclusions into bulk matrix materials, in addition to point defect scattering via alloying, was widely applied. Differential Effective Medium (DEM) method was employed to calculate two-phase heterogeneous systems. However, in most effective medium treatment, the interface scattering of matrix phonons by embedded nanoparticle was underestimated by adopting particle's projected area as scattering cross-section. Herein, modified cross-section calculations, as well as grain sizes dispersions, are applied in DEM, with the calculations then validated by comparing with Monte-Carlo simulations and existing experimental data. Predictions of lattice thermal conductivity reduction on in-situ formed Full Heusler(FH)/Half Heusler(HH) nano/matrix system are discussed.

**1. Introduction**

The conflict between growing demands of energy and limited non-renewable fossil fuel resources has drawn the world's concern over the last few decades, spurring a myriad of researchers to be dedicated in exploring clean and renewable energy, as well as improving energy transfer efficiency with conventional methods. Thermoelectric (TE) materials are of great interests because they represent a technique which can directly tap the vast reserves of currently under-used thermal energy in an environmentally friendly manner, thermoelectric generators and coolers made of such materials bear the virtues of no moving parts, no compression liquids and no noise. The performance of a TE material is evaluated by a dimensionless figure of merit $ZT(=S^2\sigma T/\kappa)$ where S is Seebeck coefficient, $\sigma$ the electrical conductivity, and $\kappa$ the thermal conductivity which consists of phonon contribution $\kappa_{ph}$ and the electronic part $\kappa_e$. While the electronic features S, $\sigma$ and $\kappa_e$ can be well tuned by doping and band structure engineering, nanostructuring techniques have also shown great promise of achieving high ZT[1-8] by significant reducing thermal conductivity, and in some cases result in Power Factor $PF(=S^2\sigma)$ enhancement.[9-10] The mechanism of nanostructuring is understandable in the sense that increased density of interfaces among embedded nanoparticles and matrix materials act as both thermal barriers and scattering centers of long-wavelength acoustic phonons[11–13] as well as energy filterers for carriers, resulting in the reduction of thermal conductivity and increase of thermal power.[3][14]

To quantitatively analyze lattice thermal conductivity reduction due to nanoinclusions, deterministic phonon Boltzmann Transport Equation(BTE),[15-18] Callaway model[19-20] derived from BTE and numerical Monte Carlo simulation,[21] with either the frequency independent gray model[12,13] or the non-gray model[14,22] are developed. There are also first-principles based calculations, using a Kubo-Greenwood style approach, accounting for disorder-induced scattering that many models which are based on the Peierls-Bolztmann equation, coherent potential approximation, or atomic models failed to take into account.[15,23] Nevertheless, even the most simplified ab-initio/first principles calculation requires extensive modeling and computing time, impeding its application in many aspects. A phenomenological approach named Effective Medium Approach (EMA) by Nan *et al*. [24,25] provides an alternative way to study two-phase heterogeneous systems. Incorporated with Average T-Matrix Approximation(ATA), Minnich and Chen[11] proposed the modified effective MFP (mean free path) for both matrix material and nanoparticles. The modification comes from phonons' single particle scattering off embedded nanoinclusions, in addition to thermal boundary resistance. However, this ATA is based on first-order T-matrix approximation, thus only applicable in small volume fractions. Poon and Limtragool extended this effective medium approach to the whole volume fraction range from 0 to 1 by introducing a Differential Effective Medium(DEM) method,[26,27] in which multiple scattering, especially at high volume fraction, was inherently implemented into DEM treatment. Experiments[2,28,29] have indicated dramatic reduction in lattice thermal conductivity $\kappa_{ph}$ even with only several percentages of nanoinclusions, which is unable to be well explained in various EMT mentioned above. The disagreement between models and experimental data strongly indicates the boundaries scattering effect was underestimated, especially at low volume faction. One possible reason for this underestimation of interfaces scattering is that scattering cross-section of embedded nanoparticle was taken as spherical particle's projected area $\pi(d/2)^2$ in previous works[11,26,27]. Herein, inspired by Kim *et al.*,[30] we implemented the much more rigorously calculated effective scattering cross section $\sigma_{eff}$, as well as grain size dispersion, into DEM simulation to achieve a better agreement with both MC simulation by Jeng[21] and experimental results.[31]

## 2. Effective scattering cross section

### 2-1. Scattering cross section

The Mie solution to Maxwell's equation[32,33] explains when electromagnetic waves encounter spherical particles, the scattering cross section varies from $\sim\omega^4$ of Rayleigh scattering to frequency independent geometrical scattering(Rayleigh-Gans-Debye scattering), as size factor $\chi=qd$ (q is the wave number, and d the particle diameter) changes from one extreme of $\chi \sim$ infinitesimal to the other as $\chi \sim$ infinity. Ying and Truell *et al.* [34,35] extended this treatment to derive the scattering cross section of phonon wave off spherical particles in a solid. Despite the relative simple form of cross sections $\sigma_{eff}$ for the two extreme cases, it is fairly difficult to establish relationships between $\sigma_{eff}$ and the various scattering parameters for a more general condition, where particle size d is comparable with the incoming wavelength $\lambda$, due to the mathematical complexity. Majumdar[36] bridged the scattering cross section of two extremes and proposed the effective cross-section $\sigma_{eff}$ for the intermediate $\chi$ to be :

$$\sigma_{eff} = \pi \left(\frac{d}{2}\right)^2 \frac{\chi^4}{\chi^4+1} \qquad (1)$$

As size factor $\chi \to 0$, thus the particle size d $\ll$ incoming wavelength $\lambda$, $\sigma_{sct} \to \pi(d/2)^2\chi^4 \sim \omega^4$, which is exactly the Rayleigh scattering; while $\chi \to \infty$, $\lambda \ll$ d, and $\sigma_{sct} \to \pi(d/2)^2$, thus the frequency independent near geometric scattering regime is obtained. This assumption is correct at describing the two extreme cases, but insufficient to depict the vibration behaviors of intermediate size factor $\chi$. Kim and Majumdar, in a later work,[30] rigorously calculated scattering cross section in two extreme regimes and suggested to bridge them together as:

$$\sigma_{total}^{-1} = \sigma_{Rayleigh}^{-1} + \sigma_{near\ geometric}^{-1} \qquad (2)$$

The effective cross section at Rayleigh regime $\sigma_{Rayleigh}$ is shown to be proportional to $\chi^4$ with the magnitude determined by mass and force difference $\Delta M/M$ and $\Delta K/K$ between embedded and matrix materials and, when $\chi$ approaches 0, $\sigma_{Rayleigh} \sim \chi^4 \to 0$, as Rayleigh limit does. The second term $\sigma_{near-geometric}$ counts for the near geometric regime scattering, and its strength is proportional to $2\pi(d/2)^2$ multiplied by an oscillating factor, which approaches 1 when $\chi$ becomes infinity. Scattering in near-geometic regime is first well studied by van de Hulst[33] for electromagnetic waves. The total effective scattering cross section $\sigma_{total}$, determined by the inverse sum of cross section at two extreme regimes, approaches 0 when $\chi$ is infinitesimal, increases to a certain value before starting to oscillate with $\chi$, finally the oscillating factor fades away and retains constant $2\pi(d/2)^2$. A more intuitive plot is well presented by Kim and Majumdar[30], and it is validated by our own calculation. It is worth noting that the scattering cross section reaches twice the projected area of spherical nanoparticles at $\chi \to \infty$, which can be understood in the sense that scattering always occurring at the edges of the nanoparticles enlarges the scattering cross section due to diffraction.[37]

To calculate the effective scattering cross section at certain particle size size parameter $\chi$, and therefore wavenumber q, is needed. Average wavenumber in host material can be estimated within Debye model by averaging over phonon spectrum of $D(\omega)*<n> \sim \omega^2/(\exp(\hbar\omega/kT)-1)$, where $D(\omega)$ is phonon density of state and $<n>$ phonon equilibrium distribution function:

$$<q> = \frac{<\omega>}{v_s} = \frac{\int_0^{\omega_d} \omega*\omega^2 * \frac{1}{e^{\frac{\hbar\omega}{kT}}-1} d\omega}{v_s \int_0^{\omega_d} \omega^2 * \frac{1}{e^{\frac{\hbar\omega}{kT}}-1} d\omega} = \frac{2}{3}\frac{\omega_D}{v_s}\frac{D_3(x)}{D_2(x)} \qquad (3)$$

where Debye frequency $\omega_d = k\theta_D/\hbar$, $v_s$ is sound speed, $x = \hbar\omega_d/kT$, and $D_n(x)$ represents n-th order of Debye function. Some calculated average wavenumbers are listed in Table I. In most realistic systems, acoustic phonons, which dominate heat transfer at temperature of interests (300~1200K), bear a characteristic wavelength λ ~ 1nm, which when compared with the nanoparticle size (generally greater than 3~5 nm, thus $\chi$>15), eventually yields $\sigma_{eff} \approx \sigma_{near-geometric}$, which is a fair enough approximation in most cases.

## 2-2. Grain size dispersion

Experiments have shown that nanoparticle sizes in most nanocomposites spread from a few nanometers to hundreds of nanometers,[23,38,39] rather than being fixed at constant value as used in models. In theory, nanoparticles with size dispersion are capable of scattering off phonons of different wavelengths over phonon spectrum, thus more effective than its analogue with constant grain sizes.[40] Jeng *et al.* [21] demonstrated that further randomness of grain size distribution doesn't help reduce lattice thermal conductivity with MC simulations, which is fairly true to some extent, however, in later simulations, changes in cross-section calculation arise if the standard deviation of grain size distribution increases, eventually lead to a considerable lower $\kappa_{ph}$. The effective scattering cross section with a normalized grain size distribution function F(x) can be expressed as:[30]

$$\sigma_{eff} = \int_{d_{min}}^{d_{max}} \sigma_{total}(x) F(x) dx \quad (4)$$

Where $d_{min}$ and $d_{max}$ are the lower and upper limit of grain size separately. Presented in Table II are 5 distribution functions and their corresponding effective scattering cross sections to be discussed in the calculations below. $F_1$ is for nanoparticles with a constant diameter, $F_2$ and $F_3$ are even distributions over different ranges, $F_4$ and $F_5$ are normalized Gamma distributions with ab = mean diameter $d_0$ and $a^{1/2}b$ = standard deviation (shape parameter a=12, scale parameter b=$d_0$/12 in $F_4$; a=3, b=$d_0$/3 in $F_5$). $F_2$ and $F_3$ are presented in order to show it is universal that grain size dispersion results in increased effective scattering cross section. In this article, Gamma distribution of $F_4$ and $F_5$ are adopted for later simulations.

The explicit form of Eqn(4) is unlikely to be derived due to the complexity of $\sigma_{total}$ depending on spherical particle diameter d, in this case resulting in a necessary numerical approach. However, there are systems of which related parameters (force constant for instance) are unavailable, a simplification can then be made to estimate the effective scattering cross section. In most realistic cases, the lower limit of obtainable grain size is around 3~5nm, with the characteristic average wavenumber q assumed to be 7.5 nm$^{-1}$(refer to table I), then $\chi$ always lies above 20~30. This falls within the region where scattering efficiency $\sigma_{sct}/\pi(d/2)^2$ weakly oscillates around 2,[30] hence $\sigma_{sct}=\sigma_{near-geometic}=\pi d^2/2$ won't result in a significant deviation from rigorous calculation from Eqn(2). Estimated effective cross section $\sigma_{eff}^*$ can then be derived from the simplified integration:

$$\sigma_{eff}^* = \int_{d_{min}}^{d_{max}} \frac{1}{2}\pi x^2 F(x)dx \quad (5)$$

The authenticity of this approximation is validated by comparing explicit integration of Eqn(5) with numerical calculation of Eqn(4) for ErAs/In$_{0.53}$Ga$_{0.47}$As system discussed by Kim *et. al*,[30] and the differences are found to be within 5%, as shown in Table II. The tiny difference between simplified and rigorous integration indicates near geometric scattering dominates in phonon scattering process at this circumstance. It is also demonstrated that increased grain size dispersion tends to result in larger effective scattering cross section, therefore reduces lattice thermal conductivity. A calculation within Callaway model has depicted similar features.[20]

**3 Validation of effective scattering cross section treatment in DEM**

Jeng *et al.* [21] has presented a gray model Monte Carlo(MC) simulation scheme to study the phonon transport and thermal conductivity in nanocomposite of Si nanoparticles hosted in Ge solid solution, the results were then compared with EMA by Nan *et al.* [24,25] and bulk Si-Ge alloy. The EMA was found to have significantly underestimated the interfaces scattering due to Si nanoparticles embedment. In a later work by Minnich and Chen,[11] as well as our earlier DEM approach,[26,27] the grain size effect was taken into count by modifying the effective MFP of both host and nanoinclusions phases. The effective MFP was obtained with Matthiessen's rule:[11]

$$\begin{cases} \frac{1}{L_{eff,host}} = \frac{1}{L_{bulk,host}} + \frac{1}{L_{sct}} \\ \frac{1}{L_{eff,np}} = \frac{1}{L_{bulk,np}} + \frac{1}{d} \end{cases} \quad (6)$$

Where $L_{sct}=V/\sigma_{eff}=\pi d^3/(6\phi\sigma_{eff})$, V is the mean volume containing one nanoparticle, d the mean diameter of spherical nanoparticles, and $\phi$ the volume faction of nanoparticles. $L_{eff,host}$ and $L_{eff,np}$ are then used to replace $L_{bulk,host}$ and $L_{bulk,np}$, therefore, corresponding lattice thermal conductivity can be written as $\kappa_{host}=c_h v_h L_{eff,host}/3$ and $\kappa_{np}=c_{np}v_{np}L_{eff,np}/3$. It is obvious that modified effective scattering cross section($\sim\pi d^2/2$) leads to a smaller $L_{sct}$, hence $L_{eff,host}$, than $\pi(d/2)^2$ as ref 11 does. Herein, we present the DEM approach with the updated scattering cross section calculated by Eqn(4), and show that the calculated results turn out to have a better fit with Jeng *et al.* gray model MC simulation.

At room temperature, most phonons are populated close to the Brillouin zone boundary where acoustic phonons' group velocities are significantly smaller than sound speed $v_s$.[41] In this case, an average group velocity is estimated by $v_g=v_s\cos(qa_0)$, where q is the wavenumber and $a_0$ is the crystal constant, over phonon spectrum in Debye model,[26,27] yielding approximately $0.38v_s$.

Considering the difficulty of obtaining series of complete parameters for a certain composite, some estimated equations are also utilized to back out unknown parameters from those already known.

Debye Temperature $\theta_D$:[42]

$$\theta_D = \frac{hv_s}{k}\left(\frac{3n}{4\pi V}\right)^{\frac{1}{3}} \tag{7}$$

Sound velocity $v_s$:

$$\begin{cases} 3v_s^{-2} = 2v_{trans}^{-2} + v_{long}^{-2} \\ \quad v_{trans} = \sqrt{G/\rho} \\ \quad v_{long} = \sqrt{E/\rho} \end{cases} \tag{8}$$

Where h is the Plank constant, k the Boltzmann constant, n the number of atoms in an unit cell, V the volume of an unit cell, G and E are shear and Yang's modulus separately, and ρ mass density.

Parameters used in our DEM calculation are listed in Table I. Presented in Figure 1 are comparisons of different simulations on Si nanoparticles ($d_0$=10nm) embedded Ge bulk matrix, It can be concluded that (1) Nan's EMA gives highest effective lattice thermal conductivity, due to lack of considering grain size effect on interfaces scattering, (2) DEM simulation with the effective scattering cross section calculated in previous sessions results in a faster drop of $\kappa_{ph}$ at low volume faction, compared with Minnich's EMA model which utilizes spherical nanoparticle's projected area $\pi d^2/4$ to count for the scattering cross section, (3) DEM simulation with the effective scattering cross section agrees well with Jeng's gray model MC simulations, which is supported by experimental results,[43,44] (4) 15% decrease in $\kappa_{ph}$ at $\phi$=0.1 is found as the standard deviation of grain size increases from 0 to 5.77nm, which strongly indicates that nanoinclusions with larger grain size dispersion is more effective in blocking thermal transfer in a solid.

## 4. Comparison and prediction

DEM with modified scattering cross section is in good agreement with MC simulations, and grain size dispersion also helps to further reduce lattice thermal conductivity of nanocomposites. Before proceeding to compare with experimental results, it is worth notifying that both EMA[11] and DEM discussed above actually deal with single particle scattering, without well considering the important multiple scattering effect among embedded nanoparticles. Indeed, as discussed by Poon and Limtragool[26], this treatment only contains 1/r! fraction of r-th order scattering, thus multiple interface scattering is underestimated. Hence, $2*\kappa_{DEM}-\kappa_{EMA}$ as suggested is utilized to represent a better approximation in properly including the 2$^{nd}$ order scattering. On the other hand, at low concentrations of nanoinclusions, phonons have little chance for multiple scattering over 2$^{nd}$ order, which validates the single particle scattering approximation at low volume fractions.

It is also worth clarifying that all the models discussed above are based on the assumption that nanoparticles are distributed evenly in solid solution of matrix materials. However, when externally mixing nanoparticles into matrix grains was performed, it always leads to aggregations of added nanoparticles at host's grain boundaries, and the sizes of aggregations vary from nm to μm. This was experimentally observed in $CoSb_3$ and $ZrO_2$ dispersed composites.[28,45] In-situ generation of nanoparticles in matrix materials via phase separation was proven to be able to effectively form evenly distributed nanoscaled second phase.[31,46]

This fact limits the calculation in this work to nanocomposites systems with in-situ grown nanoparticles.

**4-1 Comparison with PbS-PbSe nanocomposite**

Lead based chalcogenides, PbTe, PbSe and PbS, have been extensively studied by experiments over the past decades as promising TE materials, mostly due to the unique features of outstanding electrical transport properties and unusual low thermal conductivities at high temperature. In-situ phase separation process of binary systems of these Lead chalcogenides via annealing was well studied and able to be manipulated, and what is more notable is that the reduction on lattice thermal conductivity was more pronounced in the nanostructured systems compared with the solid solution analogues.[47]

Androulakis *et al.*[31] presented a systematic study on lattice thermal conductivity of PbS-PbSe binary with PbS concentration up to 16%, the observed lattice thermal conductivity at room temperature was about 15% lower than calculation based on Klemens-Drabble (KD) theory for solid solution.[48,49] The extra reduction in lattice thermal conductivity was explained by nanostructured morphology of in-situ generated PbS nanoparticles(~5nm size) embedded in PbSe solid solution. With the parameters listed in Table I, lattice thermal conductivity of PbS/PbSe nanocomposite calculated with DEM is compared with KD theory and experimental data, as depicted in Figure 2. DEM presents consistently lower lattice thermal conductivity than KD theory, indicating nanoparticle interface scattering is much more effective in reducing thermal conductivity than point defect scattering of solid solution alloying. The measured data lie between KD theory for solid solution and DEM calculation for complete nanocomposite, indicating incomplete phase separation. The latter was observed both on samples that were cooled rapidly and those post-annealed at 900K. While it is comprehensible that incomplete phase separation happens in samples which experienced rapid cooling process and to a great extent retained the solid solution at high temperatures, post-annealed process at a temperature as high as 900K might cause partial PbS to re-dissolve in the PbSe matrix. Similar temperature sensitive re-dissolving process was reported for PbS-PbTe system.[46] Based on the discussion, post-annealing at a proper temperature to allow thorough phase separation is a reasonable approach to form in-situ nanostructures which effectively scatter acoustic phonons, leading to reduction in thermal conductivity and therefore enhancement in dimensionless figure of merit ZT. Further increase of second nanoparticle phase, however, won't lead to a continuous reduction of thermal conductivity, since at certain volume fraction (in particular, >30% for PbTe/PbS system), nanoparticles of second phase will aggregate into precipitates of microscale or even larger.[46,50]

## 4-2 Prediction on $ZrNi_2Sn$/ZrNiSn Full Heusler-Half Heusler nanocomposite

Half Heusler(HH) is a well-studied series of TE materials with decent performance at high temperature[52,53], and it is also coupled with low cost and being free of toxicity. The thermoelectric performance of Half Heusler is mainly limited by its comparably high thermal conductivity. Recently, several interesting works showed improvement on Seebeck coefficient, electrical conductivity as well as reduction on lattice thermal conductivity by introducing nano-sized Full Heusler(FH) particles in HH matrix.[10,54,55] However, due to the difficulty in controlling phase separation in experiments, these publications did not show a systematic reduction in lattice thermal conductivity as the content of FH phase increases.

A prediction based on DEM calculation is presented herein to help quantitatively analyze the benefit one might gain from in-situ nano FH generation in HH matrix, as shown in Figure 3.

Matrix HH material is chosen to be ZrNiSn, and FH phase ZrNi$_2$Sn, corresponding parameters used in calculation are listed in Table I. Gamma distribution is utilized to simulate grain size dispersion with standard deviation of $0.289d_0$ and $0.577d_0$ separately, with mean grain sizes $d_0$ set to be 5nm and 10nm. Over 50% reduction on lattice thermal conductivity can be achieved by only 10% ZrNi$_2$Sn nano phases embedment with mean grain size $d_0$ of 5nm, and 40% reduction by $d_0$=10nm. Meanwhile, electrical conductivity could be enhanced since embedded metallic FH phase also acts as a carrier supplier. What is more, the Seebeck coefficient may get benefited[10,54] from the so-called energy filter effect, specifically, low energy carriers which are believed to be detrimental to Seebeck coefficient tend to be trapped at the interfaces of nanoparticles and main matrix[56]. Similar trend of reduction on lattice thermal conductivity in TiNi$_2$Sn/TiNiSn nanocomposites was reported by Birkel *et al.*[54]

## 5 Conclusion

A modified Differential Effective Medium (DEM) calculation with effective scattering cross section proposed by Kim and Majumdar was conducted to simulate the dramatic lattice thermal conductivity reduction when nano-sized particles are evenly embedded into bulk matrix materials. It was then validated by comparing with Monte-Carlo simulation results[21] of Si-Ge nanocomposite. The combined form $2*\kappa_{DEM}-\kappa_{EMA}$, which inherently includes 2$^{nd}$ order particle scattering, was recommended to replace conventional single particle scattering EMA at low volume fraction of nanoparticles. The simulation was then performed on in-situ formed nanocomposites, and the calculation was found to be in agreement with experimental results reported for PbS/PbSe system that showed phase separation.[31] Finally, prediction of lattice thermal conductivity on ZrNi$_2$Sn(FH)/ZrNiSn(HH) nanocomposite was made for a quantitative analysis.

**Acknowledgements**

The authors would like to acknowledge the financial support by a Department of Energy STTR Phase-2 Grant DE-SC0004317.

Table I - Parameters of different composites used in this work, at room temperature T=300K.

| composites | Bulk κ (W/mK) | Debye T $\theta_D$(K) | $c_p$ ($10^6$J/m$^3$K) | $v_s$ (m/s) | Avg. $v_g$(m/s)[*] | MFP (nm)[*] | $<q>_{host}$ (nm$^{-1}$)[*] | ΔM/M | ΔK/K |
|---|---|---|---|---|---|---|---|---|---|
| Si/Ge[26,38] | 150/51.7 | 645/360 | 1.66/1.67 | 6400/3900 | 2432/1482 | 110.4/64.1 | 7.62 | 0.613 | 0.301 |
| PbS/PbSe[5,31,51] | 2.4/1.9 | 145/141 | 1.58/1.44 | 2040/1910 | 775/726 | 5.88/5.45 | 6.33 | 0.164 | 0.065 |
| FH/HH[57-59] | 7.2[a]/10.1 | 318/390 | 2.29/2.03 | 2639[#]/3498 | 1003/1329 | 10.4/11.3 | 9.53 | 0.261 | 0.085 |

FH=ZrNi$_2$Sn, HH=ZrNiSn. [*]represents calculated values, [#] was calculated with Eqn(7), [a] is estimated by κ~1/ρ, this approximated relationship is derived from experimental data of ref 60-61, see supplemental material.

Table II – Spherical nanoparticles size dispersion functions dependence effective scattering cross section for ErAs/In$_{0.53}$Ga$_{0.47}$As nanocomposite with $d_0$=10nm. $F_2$ and $F_3$ are even distributions, and $F_4$ are $F_5$ are Gamma distributions with shape parameter a=12, scale parameter b=$d_0$/12 and a=3, b=$d_0$/3 separately. [(a)]numerical calc. from Eqn(2)&(4), [(b)] simplified $\sigma_{eff}^*$ from Eqn(5).

| Distribution Function F(x) | diameter size range | stdev | numerical calc.[(a)] ErAs/In$_{0.53}$Ga$_{0.47}$As $\sigma_{eff}/(\pi d^2/4)$ | Simplified[(b)] $\sigma_{eff}^*/(\pi d^2/4)$ | percentage difference (%) |
|---|---|---|---|---|---|
| $F_1(x)=\delta(x-d)$ | d | 0 | 2 | 2 | 0 |
| $F_2(x)=\dfrac{1}{d}$ | d/2~3d/2 | $\dfrac{d}{2\sqrt{3}}$ | 2.217 | 13/6 | 5 |
| $F_3(x)=\dfrac{2}{3d}$ | d/4~7d/4 | $\dfrac{d}{\sqrt{3}}$ | 2.373 | 19/8 | 0.1 |
| $F_4(x)=\dfrac{x^{a-1}e^{-\frac{x}{b}}}{b^a\Gamma(a)}$ | 0~∞ | $\dfrac{d}{2\sqrt{3}}$ | 2.189 | 13/6 | 1.0 |
| $F_5(x)=\dfrac{x^{a-1}e^{-\frac{x}{b}}}{b^a\Gamma(a)}$ | 0~∞ | $\dfrac{d}{\sqrt{3}}$ | 2.520 | 8/3 | 5.8 |

**Figure 1** Lattice thermal conductivity of Si-Ge nanocomposite dependence on Si nanoparticles' volume fraction at 300K, the nanocomposite is built up with Si nanoparticles of diameter $d_0$=10nm embedded in bulk Ge matrix.

**Figure 2** Lattice thermal conductivity of PbS/PbSe dependence on PbS nanoparticles' volume fraction at 300K, the comparisons are performed among DEM simulations for nanocomposites, experimental data and KD theory for solid solutions.

**Figure 3** Dependence of lattice thermal conductivity $\kappa_{ph}$ on $ZrNi_2Sn$ nano phase's volume fraction in $ZrNi_2Sn/ZrNiSn$ nanocomposite, at 300K. Two series of plots are presented for mean grain size $d_0$=5 nm and 10 nm separately.

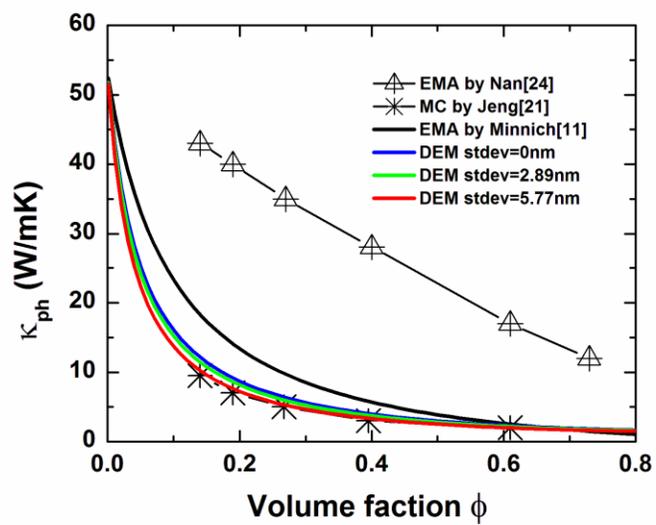

**Figure 1**

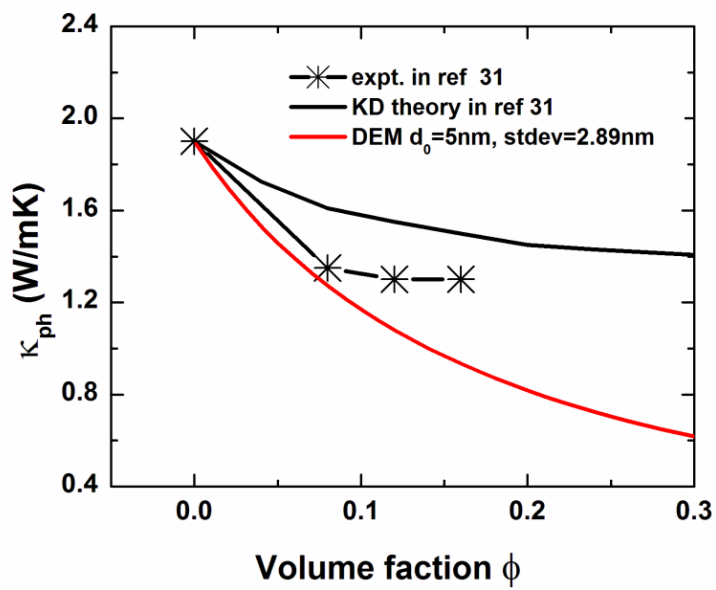

**Figure 2**

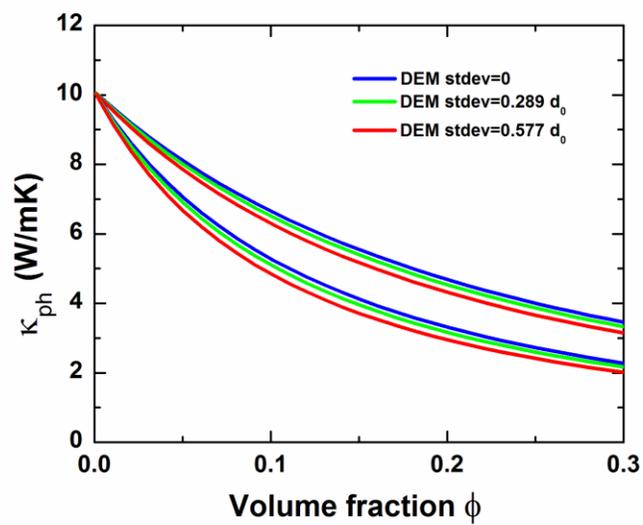

**Figure 3**

**Supplemental materials:**

Figure 4 Estimate of lattice thermal conductivity for Full Heusler ZrNi$_2$Sn, based on the linear fit of bulk lattice thermal conductivity dependence on 1/ρ. The experimental data are from ref 60-61.

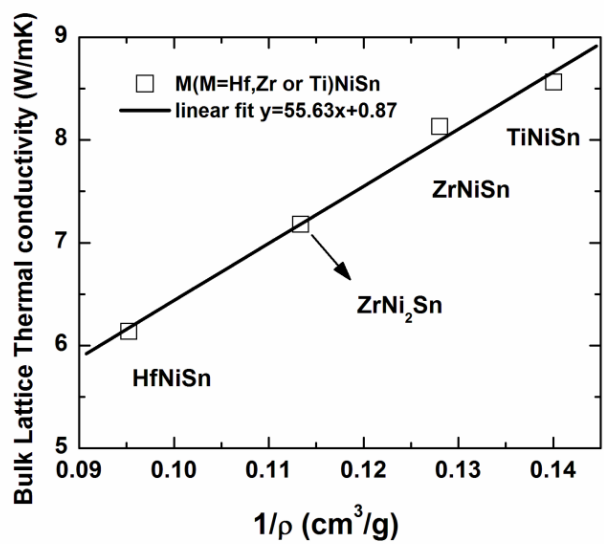